\documentclass[12pt]{article}
\usepackage{graphics}
\usepackage{psfig}
\usepackage{amsbsy}
\usepackage{a4}

\newcommand{\bt}{\beta}
\newcommand{\gm}{\gamma}

\newcommand{\ep}{\epsilon}

\newcommand{\lm}{\lambda}

\newcommand{\ph}{\phi}

\newcommand{\ps}{\psi}
\newcommand{\om}{\omega}

\newcommand{\half}{\frac{1}{2}}

\newcommand{\Tr}{\mbox{Tr}\,}

\newcommand{\psb}{\bar{\psi}}

\newcommand{\eela}[1]{\label{#1}\end{equation}}
\newcommand{\eeala}[1]{\label{#1}\end{eqnarray}}
\newcommand{\be}{\begin{equation}}
\newcommand{\ee}{\end{equation}}
\newcommand{\bea}{\begin{eqnarray}}
\newcommand{\eea}{\end{eqnarray}}
\newcommand{\bean}{\begin{eqnarray*}}
\newcommand{\eean}{\end{eqnarray*}}

\newcommand{\bra}{\langle}
\newcommand{\ket}{\rangle}

\newcommand{\HD}{{{\cal H}_D}}

\newcommand{\lesssim}{\raisebox{-.3ex}{$\enskip\stackrel{<}{\scriptstyle\sim
}\enskip$}}

\begin{document}

\title{
\vskip -100pt
{
\begin{normalsize}
\mbox{} \hfill THU-99-16\\
\mbox{} \hfill ITFA-99-16\\
\mbox{} \hfill hep-ph/9906538\\
\mbox{} \hfill June 1999\\
\vskip  70pt
\end{normalsize}
}
Particle Production and Effective Thermalization \\
in \\
Inhomogeneous Mean Field Theory
}
\author{   
Gert Aarts$^a$\thanks{email: aarts@phys.uu.nl, address after October 1, 
1999: Institut f\"ur theoretische Physik, Universit\"at 
Heidelberg, Philosophenweg 16, 69120 Heidelberg, Germany}  
\addtocounter{footnote}{1}
and
Jan Smit$^{a,b}$\thanks{email: jsmit@phys.uva.nl}\\
\normalsize
{\em $\mbox{}^a$Institute for Theoretical Physics, Utrecht University}\\
\normalsize
{\em            Princetonplein 5, 3584 CC Utrecht, the Netherlands}\\
\normalsize
{\em $\mbox{}^b$Institute for Theoretical Physics, University of
Amsterdam}\\
\normalsize
{\em            Valckenierstraat 65, 1018 XE Amsterdam, the Netherlands}
\normalsize
}
\maketitle
   
\renewcommand{\abstractname}{\normalsize Abstract} 
\begin{abstract}
\normalsize

As a toy model for dynamics in nonequilibrium quantum field theory we
consider the abelian Higgs model in $1+1$ dimensions with fermions. In the
approximate dynamical  equations, inhomogeneous classical (mean) Bose
fields are coupled to quantized fermion fields, which are treated with a
mode function expansion.  The effective equations of motion imply e.g.\
Coulomb scattering, due to the inhomogeneous gauge field.  The equations
are solved numerically.  We define time dependent fermion particle numbers
with the help of the single-time Wigner function and study particle
production starting from inhomogeneous initial conditions. The particle
numbers are compared with the Fermi-Dirac distribution parametrized by a
time dependent temperature and chemical potential. We find that the
fermions approximately thermalize locally in time. 

\end{abstract}

\newpage
 
\renewcommand{\theequation}{\arabic{section}.\arabic{equation}}
 
\section{Introduction}
\setcounter{equation}{0}
\label{secintro}

One issue that is of considerable interest in nonequilibrium quantum field
theory is equilibration and thermalization. An understanding of this is
crucial for the knowledge of e.g.\ the rate at which a quark-gluon plasma
forms in heavy ion collisions, or the rate at which a Bose gas thermalizes
locally in time during evaporative cooling.
In cosmology, the study of (pre)heating of the
universe at the end of inflation has become a topic of its own. 

Real time evolution of quantum fields can in general not be solved
exactly. A popular approach in nonequilibrium field theory is then to use 
large $N$ or Hartree-like
approximations, in which coupled equations for mean fields and their
fluctuations are solved self-consistently 
\cite{Cooper:1994hr,Kluger:1991ib,Bo:1996,Bo:1998,fermionswith1,fermionswith2,Aarts:1998td}.  
In many treatments 
available in the literature, the mean fields are taken to be homogeneous. 
Thermalization may be difficult to achieve in these approximations
\cite{Kluger:1991ib,Bo:1996,Bo:1998,fermionswith1,fermionswith2}
(see in this respect also \cite{Bettencourt:1998nf}).  
Another approach is to treat the dynamics of the low momentum modes 
classically \cite{Grigorev:1988bd,Khlebnikov:1996mc}. 

In this paper we study numerically the real time dynamics in the abelian
Higgs model in $1+1$ dimensions, extended with fermions.  The choice of
model is motivated by electroweak baryogenesis
\cite{baryo}, e.g.\ according to the scenario in
ref.\ \cite{Garcia-Bellido:1999sv}. We use a large $N$ approximation in
which the Bose fields are treated as mean fields and the fermion fields
play the role of fluctuations.  Previous numerical studies of fermions in
real time have been restricted to homogeneous mean fields, with
\cite{fermionswith1,fermionswith2} 
or without 
\cite{fermionswithout} 
back reaction.
Instead, in our case the mean fields
are inhomogeneous and they obey the full non-linear classical field 
equations  including the back reaction of the fermions. This back reaction
is represented by mode functions which obey the Dirac equation in the
presence of the mean fields.  Loosely speaking, the effective equations
describe a collection of quantum mechanical particles (the fermion modes)
coupled to classical fields (the mean fields). The particles therefore
interact and scatter amongst themselves as in classical electrodynamics. 
In particular Coulomb forces, screened by the Higgs field, are expected to
play an important role. 
 
We focus on the following dynamical problem: initially, all the energy is
contained in a few long wave lengths of the Bose fields, and the fermions
are in a vacuum state. This is a nonequilibrium situation and in course of
time energy is transferred towards the fermionic degrees of freedom,
producing fermion particles.  A motivation is partly given by inflation,
where at the end of inflation the energy stored in the inflaton field is
transferred to other degrees of freedom, leading to the (pre)heating of the
universe (see e.g.\ 
\cite{Bo:1996,fermionswith2,fermionswithout,Kofman:1997yn}). 

In the presence of spacetime dependent mean fields the identification of
particles is usually based on the concept of adiabatic particle numbers
\cite{Kluger:1991ib,Bo:1996,Bo:1998,fermionswith1,fermionswith2}. 
If the model can be described in terms of weakly coupled quasiparticles,
it should also be possible to deduce a particle number directly from
correlation functions.  In a gauge theory, simple correlation functions
such as the fermion two-point function are not gauge invariant and the
calculation would have to be performed in a fixed gauge. This introduces
an ambiguity whether the resulting particle number depends on the gauge.
However, physical observables calculated in terms of these particle
numbers should come out gauge independent. Another possibility is to use a
gauge invariant modification of the two-point function. There are several
ways in which this can be done and a simple one is by including a parallel
transporter between the two fermion fields at different points in space.
This gives a close connection to the Wigner function. 

The Wigner function has a long history in transport theory and quantum
kinetic theory
\cite{Vasak:1987um}.
It has been used to find approximative ways to deal with dynamical
issues in out-of-equilibrium plasmas.  Recently, the Wigner function has
 been used quite extensively to derive transport equations that
contain the same physics as the Hard Thermal Loops in $3+1$ dimensional
gauge plasmas 
\cite{Blaizot}. 
A comparison between so-called covariant and single-time formulations is 
given in 
\cite{singletime}. 
In this paper we use the single-time Wigner function, not for dealing with 
the dynamics but merely as a tool for identifying the particle number. We
solve the microscopic dynamics numerically in the large $N$ approximation and
then calculate the single-time Wigner function.
After averaging over space and a short interval in time to smoothen the 
effect of oscillating Bose fields, we extract the particle number
by comparing the Wigner function with that of a free fermion gas. This 
approach is obviously only valid when the theory is relatively weakly 
coupled and a description in terms of quasiparticles makes sense. 

The paper is organized as follows. In section \ref{seceom} we introduce
the model and give the effective equations of motion. A discussion of the
particle number from the fermion two-point function and the Wigner
function is given in section \ref{secwigner}.  In section
\ref{secnumerics}, we present numerical results for the real time
evolution and the time dependent particle number, starting from a
nonequilibrium initial state.  To see whether the particles are
effectively thermalized, we compare the particle numbers with the
Fermi-Dirac distribution depending on time dependent temperature and
chemical potential. The results are summarized in the Conclusion. The
equations are solved numerically on a lattice in space and time. Some
aspects of the lattice implementation are briefly described in appendix
\ref{seclat}.  Details relating the Wigner function and the gauge fixed
two-point function are given in appendix \ref{app}. 

We refer to our previous paper \cite{Aarts:1998td} for more details on the
model, the effective equations of motion, and a discussion of the lattice
implementation.

\section{Action and mean field equations of motion}
\setcounter{equation}{0}
\label{seceom}
We consider the abelian Higgs model in $1+1$ dimensions, with 
$n_f=2$ flavours of fermions. The action reads 
\bean
 S \!\!\!&=&\!\!\! -\int d^2 x  \Bigg[ 
\frac{1}{4e^2}F_{\mu\nu}F^{\mu\nu} +
 (D_\mu \phi)^*D^\mu \phi + 
  \lambda\left(\phi^*\phi - v^2/2\right)^2\\
&&\;\;\;\;\;\;\;+ 
\bar\psi_i\gm^\mu(\partial_\mu - iqA_\mu)\psi_i 
+ \half \sum_i (G_i\psi_i^T{\cal C}^\dagger\phi^*\psi_i 
- G_i\bar\psi_i{\cal C}\phi\bar\psi_i^T)\Bigg].
\eean
Here $F_{\mu\nu} = \partial_\mu A_\nu-\partial_\nu A_\mu$, $D_\mu =
\partial_\mu - iA_\mu$, ${\cal C}$ denotes the charge conjugation matrix
and $\bar\psi = \psi^\dagger\beta$. 
An explicit representation for the gamma-matrices is $\gm^1=\sigma_1, 
{\cal C} = \beta = i\gm^0=\sigma_2$, and $\gm_5=-\gm^0\gm^1=\sigma_3$.
Space is a circle with circumference $L$ and
the Bose (fermion) fields obey (anti)periodic boundary conditions.
Other conventions can be found in
\cite{Aarts:1998td}.  The choice of this model, in particular the form of
the Yukawa coupling, is motivated by electroweak baryogenesis, but we will
not elaborate on those aspects here (see instead \cite{Aarts:1998td}). The
fermion field carries a flavour index $i=1,2=n_f$
and the 
Yukawa coupling $G_i$ depends on the flavour: 
$G_1=-G_2\equiv G$. Local gauge symmetry acts as $\phi\to e^{i\xi}\phi,
\psi_i\to e^{iq\xi}\psi_i, A_\mu\to A_\mu +\partial_\mu \xi$, with
$q=\half$, so that the Yukawa term is gauge invariant. In this paper we
restrict ourselves, however, to massless fermions and put the Yukawa term
to zero, $G=0$. 

The global symmetry $\psi_i \to e^{i\om\gm_5}\psi_i$ is broken
in the quantum theory due to the axial anomaly, and
\[
\partial_\mu j_5^\mu = n_f\frac{1}{4\pi}\ep^{\mu\nu}F_{\mu\nu} = 
n_f\partial_\mu C^\mu,
\]
with the axial current $j_5^\mu = i\bar\psi_i\gm^\mu\gm_5\psi_i$ 
and the Chern-Simons current $C^\mu = \ep^{\mu\nu}A_{\nu}/2\pi$.
A change in the axial charge is proportional to a change in the Chern-Simons 
number \be
\label{eqCQ5}  
Q_5(t)-Q_5(0)= n_f(C(t)-C(0)),
\ee
with $Q_5(t) = \int dx\,j^0_5(x,t)$, and $C(t)=-(2\pi)^{-1}\int 
dx\,A_1(x,t)$. As in the electroweak theory, integer values of the 
Chern-Simons number label the non-trivial vacua of the bosonic theory. 
The $q=\half$ charged fermions change the periodicity of the classical 
bosonic ground state: only vacua with Chern-Simons numbers that differ an 
even integer (instead of any integer) are connected by large gauge 
transformations.
The vacua are separated by finite energy barriers, the sphaleron 
configuration at half integer $C$.

As an approximation to solve the dynamics, we treat the Bose fields as
mean fields and the fermion fields as fluctuations. A formal derivation
can be given by duplicating the fermion fields $N$ times and taking the 
large $N$ limit, after a proper rescaling of the coupling constants and
fields \cite{Cooper:1994hr,Aarts:1998td}. The resulting equations are 
(in the $A_0=0$ gauge) 
\bea
\label{eqAeq}
\partial^2_0A_{1}(x,t) \!\!\!&=&\!\!\! e^2 \left[j_{h}^1(x,t) + \bra
j_{f}^1(x,t)\ket\right],\\
\label{eqhiggs}
\partial_0^2\phi(x,t) \!\!\!&=&\!\!\! D^2_1\phi(x,t) - 2\lambda
\left[|\phi(x,t)|^2 -  v^2/2\right]\phi(x,t),
\eea
together with Gauss' law
\be
\label{eqgauss}
\partial_1\partial_0 A_{1}(x,t) = -e^2 \left[j_{h}^0(x,t) + \bra
j_{f}^0(x,t)\ket\right].
\ee
The scalar contribution to the current, $j^\mu_h$, is given by
\be
\label{eqhiggscurrent}
 j^\mu_{h}(x,t) =  i(D^\mu\phi(x,t))^*\phi(x,t)
-i\phi^*(x,t)D^\mu\phi(x,t).
\ee
The fields $A_1$ and $\phi$ represent mean fields, which  can be 
inhomogeneous in general,
\[ A_1(x,t)\equiv\bra 
A_1(x,t)\ket,\;\;\;\; \phi(x,t)\equiv\bra\phi(x,t)\ket.
\]
The fermion contribution to the current is
\[
 j^\mu_{f}(x,t) = \frac{q}{2}i[\bar\psi_i(x,t),\gm^\mu\psi_i(x,t)], 
\]
and its expectation value represents the fermion back reaction.
The commutator ensures that the current is odd under charge conjugation.

The most demanding part of this approach is the calculation of the back 
reaction. In the case of homogeneous mean fields, the usual manner is to 
use a mode function expansion. We use the same method in the  
inhomogeneous case as well.

The back reaction is calculated by expanding the fermion 
fields in a complete set of eigenspinors of the initial Dirac hamiltonian 
at $t=0$. This leads to a mode function expansion, 
\be 
\label{eqexp}
\psi_i(x,t) = \sum_p \left[ b_{pi}u_{pi}(x,t) +
d^\dagger_{pi}v_{pi}(x,t)\right]. 
\ee 
The coefficients
$b^{(\dagger)}_{pi}, d^{(\dagger)}_{pi}$ are time independent and
represent the annihilation (creation) operators of particles resp.\
antiparticles at the initial time. They obey the usual 
anticommutation relations 
\[
\{b^\dagger_{pi},b_{p'i'}\} = \delta_{pp'}\delta_{ii'},
\;\;\;\; 
 \{d^\dagger_{pi},d_{p'i'}\} = \delta_{pp'}\delta_{ii'}, 
\] 
and zero
for others. 
The expectation values of the creation and
annihilation operators in (\ref{eqexp}) determines the initial state for 
the fermions. We take vacuum initial conditions for the fermions, so that 
the only non-zero expectation values are 
\be
\label{eqvac}
\bra b_{pi}b^\dagger_{pi}\ket = \bra
d_{pi}d^\dagger_{pi}\ket = 1.
\ee
The fermion contribution to the current can now be expressed in terms of 
the mode functions, 
\be 
\label{eqcumode}
\bra j^\mu_f(x,t)\ket = -
\sum_{p}\frac{q}{2}\left[ \bar u_{pi}(x,t)i\gm^\mu u_{pi}(x,t) - \bar
v_{pi}(x,t)i\gm^\mu v_{pi}(x,t)\right]. \ee
The time dependence is carried by the spinor mode functions
$u$ and $v$, which are solutions of the Dirac equation in the presence of
the gauge field,
\be 
\label{eqmode} 
i\partial_0 u_{pi}(x,t) =
\HD[A_1(x,t)]u_{pi}(x,t),
\ee 
and similar for $v_{pi}$. 
The Dirac hamiltonian has the usual form 
\be
\label{eqHD} 
\HD = -i\alpha^1(\partial_1-iqA_1), \;\;\;\;\alpha^1=-\gm^0\gm^1.
\ee
The initial conditions for the mode functions are given by 
\[ 
u_{pi}(x,0) = \frac{1}{\sqrt{L}}e^{ipx}u_{pi},\;\;\;\; 
v_{pi}(x,0) = \frac{1}{\sqrt{L}}e^{-ipx}v_{pi},
\] 
with $u_{pi}(v_{pi})$ positive (negative) energy spinors of the Dirac 
hamiltonian at $t=0$: $p$ labels the momentum eigenstates of the 
initial Dirac hamiltonian. It takes the values
\[ 
p=\frac{2\pi}{L}(n-\half),
\]
due to the antiperiodic boundary conditions on the fermion fields.
For these initial conditions with vacuum 
expectation values (\ref{eqvac}), the current (\ref{eqcumode}) vanishes at 
$t=0$. 

If the mean fields are restricted to be homogeneous, plane waves $e^{\pm
ipx}$ multiplying the mode functions can be factored out at all times, and
each mode function is the solution of an {\em ordinary} differential
equation. In the more general case that the mean fields are allowed to be
inhomogeneous, this is not possible, and a large set of {\em partial}
differential equations has to be solved. In that case the label $p$ looses
its interpretation for $t>0$, and it simply labels a complete set. 

The equations of motion are free from ultraviolet divergences and the sum
over the mode functions in (\ref{eqcumode}) is finite. In the higher
dimensional case charge renormalization would be necessary 
\cite{Cooper:1994hr}. For a study of
renormalization in the presence of the Yukawa term we refer to our
previous paper \cite{Aarts:1998td}. 
The energy of the system is the sum of the energy in the Bose fields and
the fermion fields. The energy of the fermions at time $t$ is given by the
expectation value $E_F(t)=\sum_{x}\bra\psi_i^\dagger(x,t)\HD
\psi_i(x,t)\ket$ and can be expressed as a sum over mode functions. This
sum is quadratically divergent. 
In the numerical implementation this is regulated by the lattice cutoff: 
the effective equations are solved using a discretization on a lattice in
space and real time (see appendix \ref{seclat}). We renormalize the energy
by subtracting the bare fermion energy at $t=0$.

\section{Particle number and the Wigner function}
\setcounter{equation}{0}
\label{secwigner}

Observables which may give insight about what happens on the
microscopic level during time evolution are current, charge and energy
densities.  These are gauge invariant quantities whose physical meaning is
clear and unambiguous and they are straightforward to calculate.  In
addition to this, it has been shown in the case of homogeneous mean fields
that more insight can be found with the concept of adiabatic or
instantaneous particle number. 

The adiabatic particle number is based on an expansion of 
the (interacting) quantum field at each instant in time in terms of a 
complete set of eigenfunctions of the Dirac hamiltonian at time $t$
(we suppress flavour indices here) 
\be 
\label{eqexp2}
\psi(x,t) = \sum_{E} \left[ \tilde b_{E}(t)\tilde u_{E}(x,t) + \tilde 
d_{E}^\dagger(t)\tilde v_{E}(x,t)\right].
\ee
The coefficients of these eigenfunctions are then identified  
with time dependent creation and annihilation operators. 
In (\ref{eqexp2}) the eigenfunctions are denoted with $\tilde 
u_{E}, \tilde v_{E}$, and the complete set is labeled with 
$E$.
A Bogoliubov transformation relates the expansions (\ref{eqexp}) and 
(\ref{eqexp2}).
The instantaneous particle numbers at time $t$ are then {\em 
defined} by the expectation values
\[
N_{E}(t) = \bra \tilde b^\dagger_{E}(t)\tilde b_{E}(t)\ket,\;\;\;\;
\bar N_{E}(t) = \bra \tilde d^\dagger_{E}(t)\tilde 
d_{E}(t)\ket.
\]
In the case of homogeneous mean fields the instantaneous eigenfunctions
can be calculated analytically, in terms of plane waves. For the
inhomogeneous case the diagonalization will in general have to be
performed numerically, which makes the calculation rather involved.
Furthermore, the instantaneous eigenmodes of the Dirac hamiltonian are not
easily interpreted in terms of familiar concepts like quasiparticles,
e.g.\ concerning the relation between $E$ and the quasiparticle momentum.
Therefore we now discuss another approach, 
using correlation functions.\footnote{
Correlations functions have been used in other contexts as well:  
see e.g.\ \cite{Ibaceta:1998} for a study of defect formation during 
nonequilibrium evolution with correlation functions.}

The identification of particles 
can be motivated in general (for a weakly coupled system) by comparison with
the noninteracting case, and for free fields the two-point function
provides all the available information. 
Fields in equilibrium are homogeneous in space and time.
In the interacting situation, one 
can look for subsets of the system which are effectively in equilibrium.
Such subsets will only exist {\em locally in space and time}, 
and one way to enforce homogeneity is averaging
over small subset-volumes of spacetime. This then leads to 
a definition of particle numbers locally in space and time. 
The optimal size of the subsets is not known a priori, and this may  
introduce a certain ambiguity. The dependence on the size of the subsets 
can be studied by changing it.

One way to implement this idea is by studying the equal-time
two-point function 
\be 
\label{eqfermionprop} 
S(x,y;t) =
\langle\ps(x,t)\psb(y,t)\rangle, 
\ee
in particular in the following manner
\be 
\label{Sdelta} S_{\rm av}(p,t) = \frac{1}{t_{\rm av}}\int_{t-t_{\rm
av}/2}^{t+t_{\rm av}/2} dt'\, S(p,t'), \;\;\;\; S(p,t) = \int
dz\,e^{-ipz}\frac{1}{L}\int dx\, S(x,x+z;t). 
\ee 
The time average over a short time interval $t_{\rm av}$ will 
smoothen oscillations in the Bose fields, 
and instead of averaging over smaller spatial subsets, we choose here to  
integrate over the complete spatial volume.\footnote{
In equilibrium the two-point function in (\ref{Sdelta}) does not
depend on $x$ at all because of translational invariance. Slightly out of
equilibrium, the long wave length inhomogeneities around equilibrium in
the two-point function (or rather in the Wigner function, see below) can
be used to derive transport or kinetic equations by applying a gradient
expansion \cite{Vasak:1987um,Blaizot}.} Hence, the particle numbers will be
defined locally in time only. This simplifies the analysis and has the
positive effect of diminishing the fluctuations.
As a side remark, it can be checked explicitly that this approach reduces 
to a commonly used definition of particle number for the case of homogeneous
mean fields \cite{fermionswith1,fermionswith2}.

The fermion correlator is not gauge invariant, but the concept of
quasiparticles is not gauge invariant in the first place, although using
them to compute observables such as energy and pressure should give gauge
invariant answers. 
Therefore, before averaging over space and time we need to fix the gauge. 
A suitable choice is the temporal Coulomb 
gauge, $A_0 = 0$, $\partial_1 A_1 = 0$, in which the remaining freedom of 
`large' gauge transformations is removed by requiring $C= -LA_1/(2\pi)\in 
(-1,1]$.  

There are of course other possibilities. 
One is to use gauge invariant modifications of the fermion 
correlator (\ref{eqfermionprop}), for example,  
\be
\label{eqW} 
W(x,y;t) = \bra\psi(x,t)\bar{\psi}(y,t)U(y,x;t)\ket,
\ee
where $U$ is the parallel transporter 
\[ 
U(y,x;t) = \exp \left[ -iq\int_y^x ds\, A_1(s,t)\right].
\]
The path is along the shortest straight line connecting $x$ and $y$. 
Other paths are possible as well, which leads to a class of gauge 
invariant modifications of the two-point function (\ref{eqfermionprop}). 
The arbitrariness in choosing a gauge is now replaced by choosing a gauge 
invariant correlator. 
The above $W(x,y;t)$ is closely related to the single-time Wigner 
function 
\cite{Vasak:1987um,singletime}.
In terms of the relative coordinate $z=x-y$ and the center-of-mass 
coordinate $X=\half(x+y)$, the single-time Wigner function
is given by a spatial Fourier transform of (\ref{eqW}) with respect to $z$,
\[
W(X,p;t) = \int 
dz\, e^{-ipz}\, W(X+\half z, X-\half z;t).
\]
We shall use its spatial and temporal average,
\be
\label{eqWav}
W_{\rm av}(p,t) =
\frac{1}{t_{\rm av}}\int_{t-t_{\rm av}/2}^{t+t_{\rm av}/2} dt'\, W(p,t'),
\;\;\;\;
W(p,t) = \frac{1}{L}\int dX\, W(X,p;t),
\ee
similar as for the two-point function, (\ref{Sdelta}).
In appendix \ref{app} we show that the correlators $S(p,t)$ and 
$W(p,t)$ are actually closely related by mere interpolation
between the discrete momenta $p$. In the following
we shall continue with the Wigner function.

In the spinor decomposition (in $1+1$ dimensions), four functions appear
\be
\label{eqWdecomp}
W(X,p;t) =  {\cal F}(X,p;t) + i\gm^\mu{\cal V}_\mu(X,p;t) + i\gm_5{\cal 
P}(X,p;t), \ee
that are real, due to the elementary property $W^\dagger(X,p;t) = 
\beta W(X,p;t)\beta$.
We shall use the obvious notation ${\cal F}_{\rm av}(p,t)$, etc., for 
temporally and spatially averaged coefficients.

For fermions characterized by momentum $p$ and mass $m$, the free field 
expression for the Wigner function is given by
\bea
\label{eqWfree}
W_{\rm free}(X,p) \!\!\!&=&\!\!\!
(1-N_p-\bar N_{-p})\frac{m-i\gm^1p}{2E_p} + 
\half i\gm^0(1-N_p+\bar N_{-p}),\\ 
\nonumber
E_p \!\!\!&=&\!\!\! \sqrt{m^2 + p^2}.
\eea
The arbitrary (anti)particle occupation numbers are given by
$N_p$ ($\bar N_p$).
Since this is independent of $X$, we now take as a 
possible definition of time dependent particle numbers $N_p(t)$, $\bar 
N_p(t)$ and effective mass 
$m_p(t)$, the solution of the three equations
\bea
\label{eqF}
{\cal F}_{\rm av}(p,t) \!\!\!&=&\!\!\! \half\Tr W_{\rm av}(p,t) 
= \half\left[1-N_p(t) - \bar N_{-p}(t)\right]\frac{m_p(t)}{\sqrt{m_p^2(t) + 
p^2}}, \\
\label{eqV1}
-{\cal V}_{1{\rm av}}(p,t) \!\!\!&=&\!\!\! \half\Tr i\gm^1 W_{{\rm av}}(p,t) 
= \half\left[1-N_p(t) - \bar N_{-p}(t)\right] \frac{p}{\sqrt{m_p^2(t) + 
p^2}}, \\
\label{eqV0}
{\cal V}_{0{\rm av}}(p,t) \!\!\!&=&\!\!\! \half\Tr i\gm^0 W_{{\rm av}}(p,t) 
= \half\left[1-N_p(t) + \bar N_{-p}(t)\right].
\eea
The explicit solution of these equations is given in appendix \ref{app}. 
We have added a subscript $p$ on the effective mass parameter to 
indicate a possible momentum dependence. 
The interpretation of $m_p$ as a 
mass is of course best when $m_p(t)$ is momentum independent,
as in the free field case.
A comparison with free Wigner function (\ref{eqWfree}) gives zero for 
the fourth function ${\cal P}_{\rm av}(p,t) = -\Tr i\gm_5 W_{{\rm av}}(p,t)$. 
Hence, this coefficient cannot be used to find new information on the 
particle numbers or effective mass, but it serves as a consistency check.

Note that when the mean fields are homogeneous, particles and
antiparticles with momentum $p$ can only be produced in pairs, and
$N_p=\bar N_{-p}$. In general, they can be different. It should be clear
that we use the Wigner function only to identify the instantaneous
particle number at time $t$ (i.e.\ after averaging over the short interval
$t_{\rm av}$). Therefore we consider the single-time Wigner function
instead of the covariant one. We do not attempt to
solve the dynamics using the Wigner function approach.  

Our goal is now to solve the effective equations for the Bose fields and 
the spinor mode functions.
The Wigner function can be expressed in terms  of the mode functions 
using the expansion (\ref{eqexp}), and from a calculation of ${\cal 
F}_{\rm av}(p,t)$, ${\cal V}_{\mu{\rm av}}(p,t)$ and ${\cal P}_{\rm av}(p,t)$
the particle number can be extracted. 
This will be the subject of the following section.

\section{Particle production}
\setcounter{equation}{0}
\label{secnumerics}

We solve the closed set of effective equations
(\ref{eqAeq}--\ref{eqhiggscurrent}, \ref{eqcumode}--\ref{eqHD})
numerically. 
In this paper we take the following parameters: the dimensionful
parameters are related by $\lambda/e^2=0.25, eL=3.2$ and the dimensionless 
parameter in the scalar potential is $v^2=8$. 
Then the volume is not very large but also not too small: in terms of the 
tree level boson masses,
$m_{\ph} L = \sqrt{2\lm}vL \approx 6.4$, $m_A L = evL \approx 9.1$. 
Furthermore, the couplings are fairly weak: $e^2/m^2_{\ph} = 0.25$.
Some details on the lattice implementation and choice of
parameters can be found in appendix \ref{seclat}. 

The inhomogeneous Bose fields are initialized such that all the energy
initially resides in the long wave lengths. We take vacuum field
configurations $\phi(x,0)=v/\sqrt{2}$, $A_1(x,0)=0$, and use space dependent
momenta.
The results presented below are obtained starting with
\[
\partial_0\phi(x, 0) = 3e\cos \frac{2\pi x}{L} +2ie\cos \frac{2\pi x}{L},
\;\;\;\; \partial_0A_1(x,0)=\mbox{Gauss' law} + e^2.
\]
The electric field $\partial_0A_1(x,0)$ is determined almost completely 
by the Gauss law (\ref{eqgauss}), up to a constant. 

\begin{figure}[ht]
\centerline{\psfig{figure=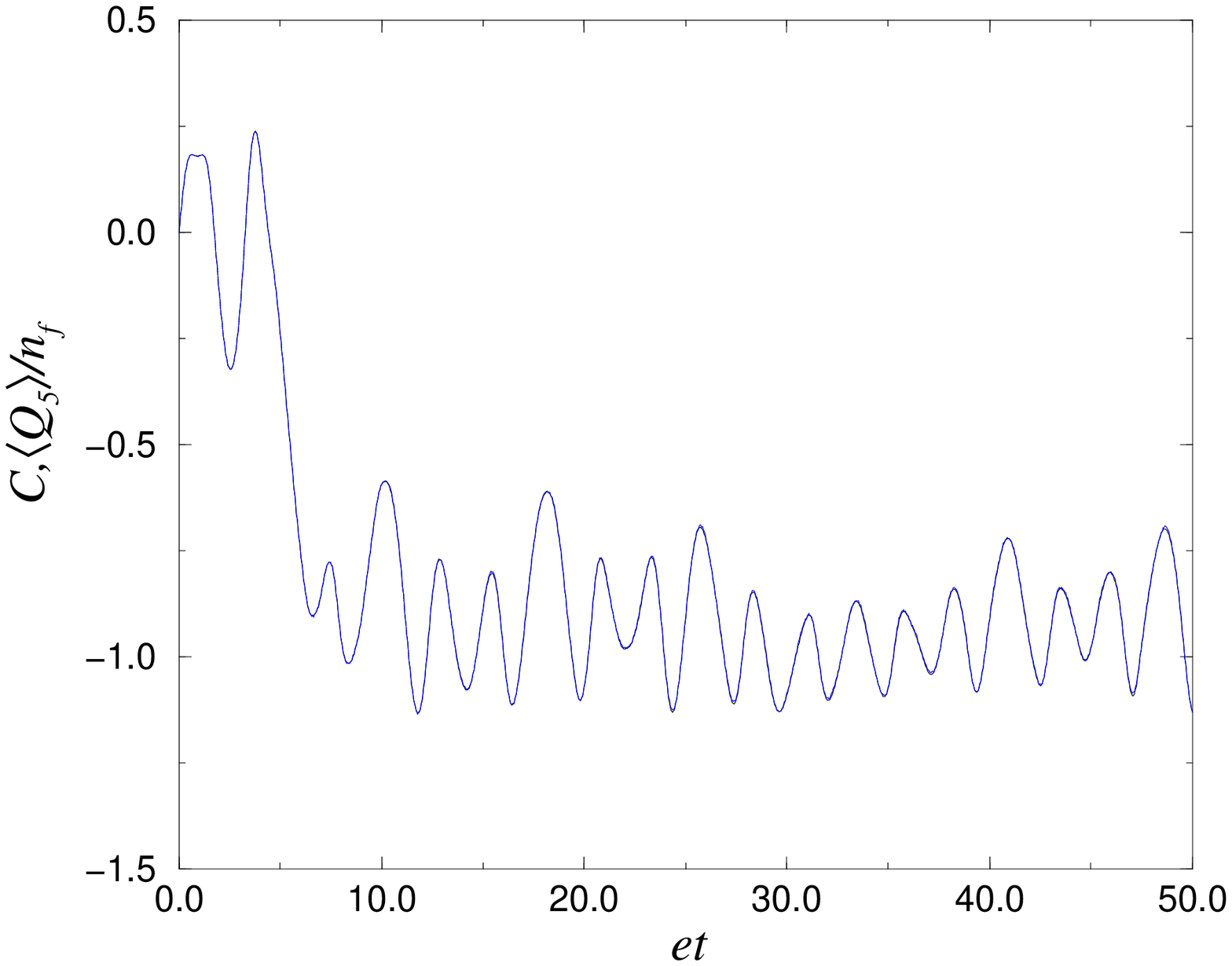,width=16.0cm}}
\caption{Chern-Simons number $C$ and axial charge per flavour $\bra 
Q_5\ket/n_f$ 
versus $et$. The lines fall on top of each other, as expected from the 
anomaly equation.
} \label{figC}
\end{figure}

\begin{figure}
\centerline{\psfig{figure=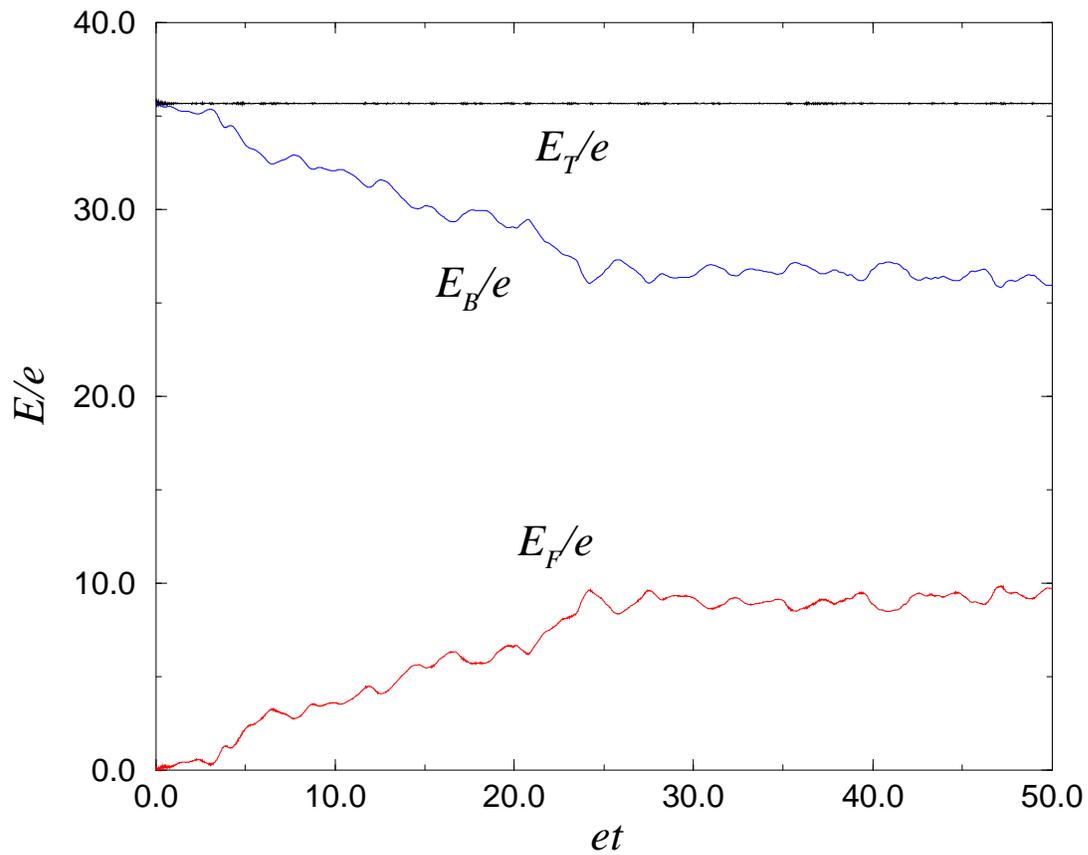,width=16.0cm}}
\caption{Energy in units of $e$ in the Bose fields ($E_B$), fermion fields 
($E_F$) and their conserved sum $E_T$ versus $et$. }
\label{figenergy}
\end{figure}

The Chern-Simons number $C$ and the axial charge per flavour $\bra 
Q_5\ket/n_f$, which agree in
accordance with the anomaly equation (\ref{eqCQ5}), are shown during time
evolution in fig.\ \ref{figC}. Time is given in units of $1/e$, which has 
the dimension of mass in $1+1$ dimensions.
We see one sphaleron transition. The
energy of the Bose fields, the fermion fields and the total conserved
energy are shown in fig.\ \ref{figenergy}. Initially, the (renormalized)
fermion energy is zero, because of the vacuum initial conditions for the
fermions. 
We see that there is energy transfer towards the fermion degrees of 
freedom. 

To analyse how the energy is distributed in the fermion subsystem we have
calculated the four functions ${\cal F}(p,t), {\cal V}_\mu(p,t)$ and
${\cal P}(p,t)$ which appear in the spinor decomposition of the Wigner
function (\ref{eqWav}, \ref{eqWdecomp}),\footnote{The results for the
particle numbers presented below are normalized for one flavour, i.e.\
divided by $n_f=2$.} and averaged them over a relatively short time
interval $et_{\rm av}=4$.  This interval is comparable with the
elementary periods present in the Bose system: 
$et_A=2\pi e/m_A\approx 2.2, et_\phi\approx 3.1$. 
Below, the time $et$ refers to the center of the interval $et_{\rm av}$. 
With 
regard to the spatial momentum dependence, $p$ is written in terms of
the dimensionless combination $ap$, where $a$ is the lattice spacing. Note 
that the continuum regime is where the momentum $p$ is small with respect 
to the cutoff $\pi/a$, e.g.\ $|ap| \lesssim 0.5$ (see appendix \ref{seclat}).

\begin{figure}
\centerline{\psfig{figure=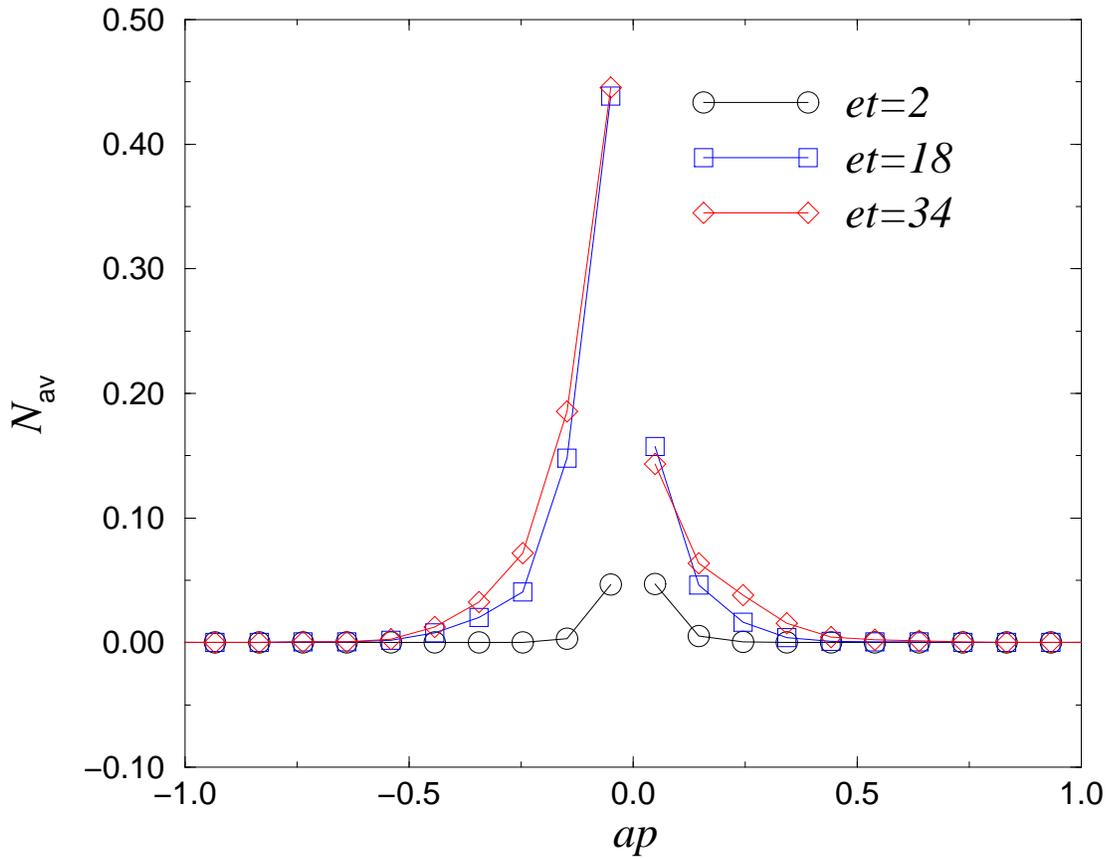,width=16.0cm}}
\caption{Time dependent particle number $N_{\rm av}$ versus $ap$, the 
momentum in lattice units, at three different times. The particle numbers 
are obtained from a time average over an interval of length 
$et_{\rm av}=4$ and $et$ denotes the center of the interval.}
\label{figN} 
\end{figure}

We solve the time dependent particle number from the three equations that
are suggested by the free Wigner function (\ref{eqF}--\ref{eqV0}), in the
combination 
\[ 
N_{{\rm av}p}(t) \equiv \half\left(N_p(t)+\bar N_{-p}(t)\right). 
\] 
The result for the averaged particle number is shown in fig.\ \ref{figN}.
The particle numbers of the modes with $|ap|>1$ are consistent with zero,
up to numerical precision. These modes are not excited at all.  As can be
seen from fig.\ \ref{figN}, essentially only the physical (i.e.\ without
lattice artefacts) fermions (with $|ap| \lesssim 0.5$) are excited.
Furthermore the distributions are much smoother than those obtained with
homogeneous mean fields \cite{fermionswith1,fermionswith2}. 
Note that the distribution functions are not symmetric under $p\to-p$. 
This will be discussed below.

With respect to the time dependent mass $m_p(t)$ in (\ref{eqF}--\ref{eqV0}),
we find no significant deviation from the free dispersion relation
$E=|p|$.\footnote{In the analysis we use of course the lattice dispersion
relation.} Finally, we find that ${\cal V}_{0{\rm av}}(p,t)-1/2$ and
${\cal P}_{\rm av}(p,t)$ are consistent with zero for momenta $|ap|>0.5$. 
For smaller momenta they fluctuate around zero. The amplitude in the 
case of ${\cal P}_{\rm av}(p,t)$ is smaller than $0.005$.

In order to see whether the fermion subsystem effectively thermalizes 
locally in time we compare it with the Fermi-Dirac distribution 
function, depending on a time dependent temperature $T(t)=1/\beta(t)$ 
and chemical potential $\mu(t)$,
\be 
f_p(t) = \frac{1}{\exp[\bt(t)(E_p-\mu(t) q_{5p})]+1}.
\ee
The chemical potential is coupled to the axial charge density, and 
we use the notation $q_{5p}=\mbox{sign}(p)=\mbox{sign}(\gm_5)$ to denote the 
chirality. Note that this term breaks the symmetry $p\to-p$.
In thermal and chemical equilibrium $N_p(t) = \bar{N}_{-p}(t) =  N_{{\rm 
av}p}(t) = f_p(t)$.

\begin{figure}
\centerline{\psfig{figure=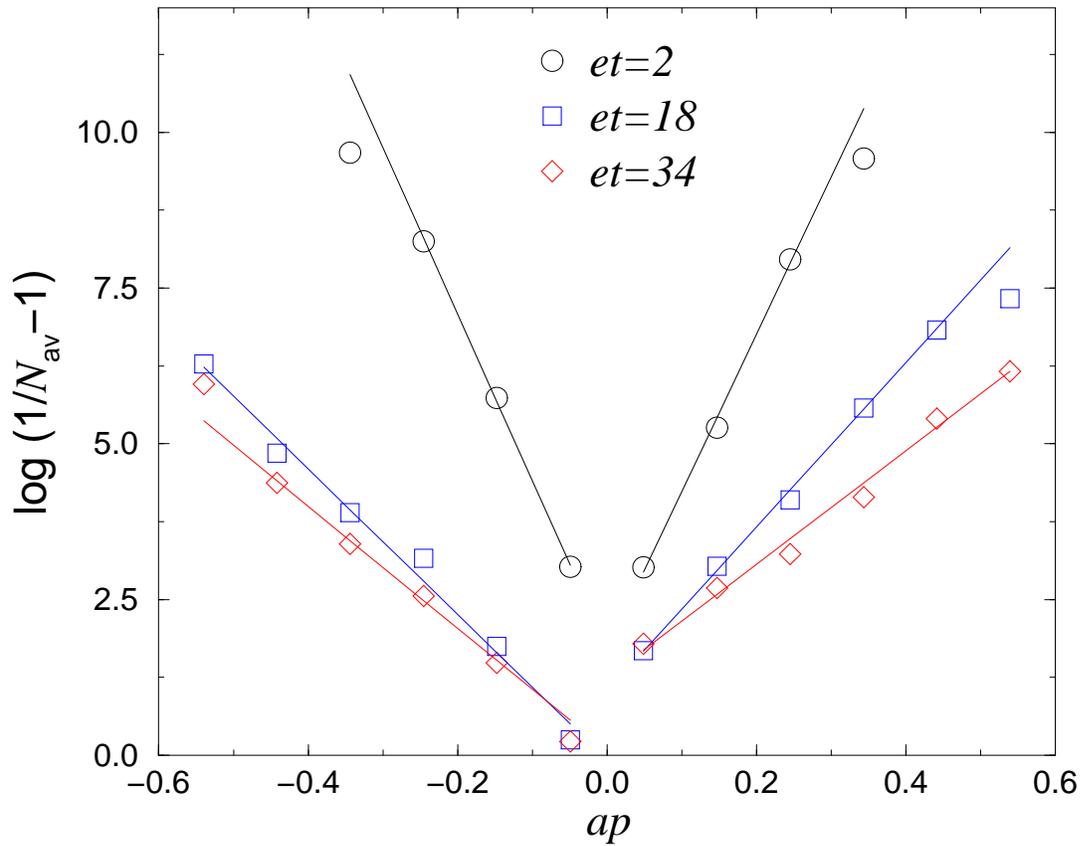,width=16.0cm}}
\caption{Transformed particle number $\log(N^{-1}_{\rm av}-1)$ for the 
data of fig.\ \ref{figN}. In equilibrium this would give $\bt(t)[|p| 
\pm\mu(t)]$. The lines are obtained from a straight line fit.  
} \label{figstraight}
\end{figure}

To compare the particle numbers with the Fermi-Dirac distribution, we 
plot $\log (N^{-1}_{\rm av}-1)$ versus $ap$. In equilibrium this would 
result in a straight line $\bt(t)[|p| \pm\mu(t)]$. The results are shown in
fig.\ \ref{figstraight}. We see that the points seem to lie approximately 
on straight lines, with small deviations present. The lines 
through the points are obtained from a straight line fit.
Note that the data denoted with $et=2$ represent the averaged particle 
number in the first interval, $0<et<4$, which contains less than 1.5 
oscillations of the Chern-Simons number (see fig.\ \ref{figC}). Already 
here the low momentum modes 
show approximate thermalization.

\begin{figure}
\centerline{\psfig{figure=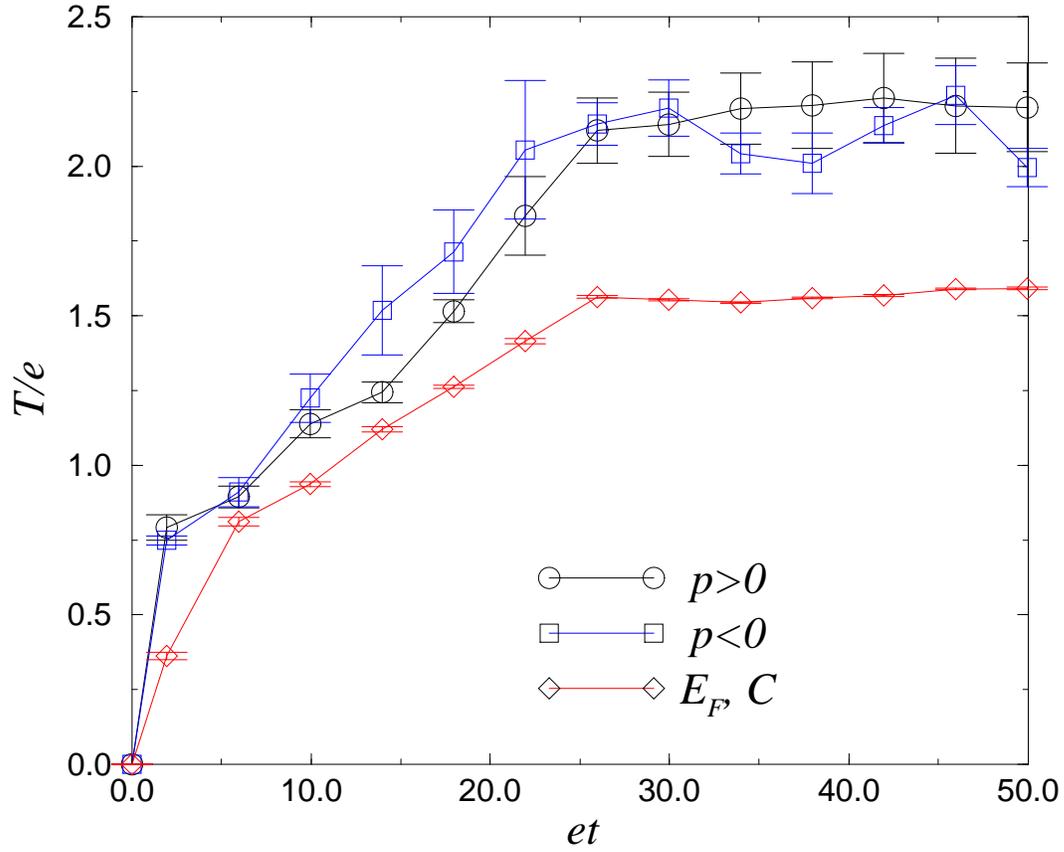,width=16.0cm}}
\caption{Time dependent effective temperature versus $et$, obtained from 
least square fits (with errors) from the transformed particle numbers as 
in fig.\ \ref{figstraight}, for $p>0, p<0$ separately. The line denoted 
with $E_F, C$ is obtained from (\ref{eqenergy}), i.e.\ 
{\em assuming}\, that the fermions are in thermal and chemical equilibrium.}
\label{figT}
\end{figure}

\begin{figure}
\centerline{\psfig{figure=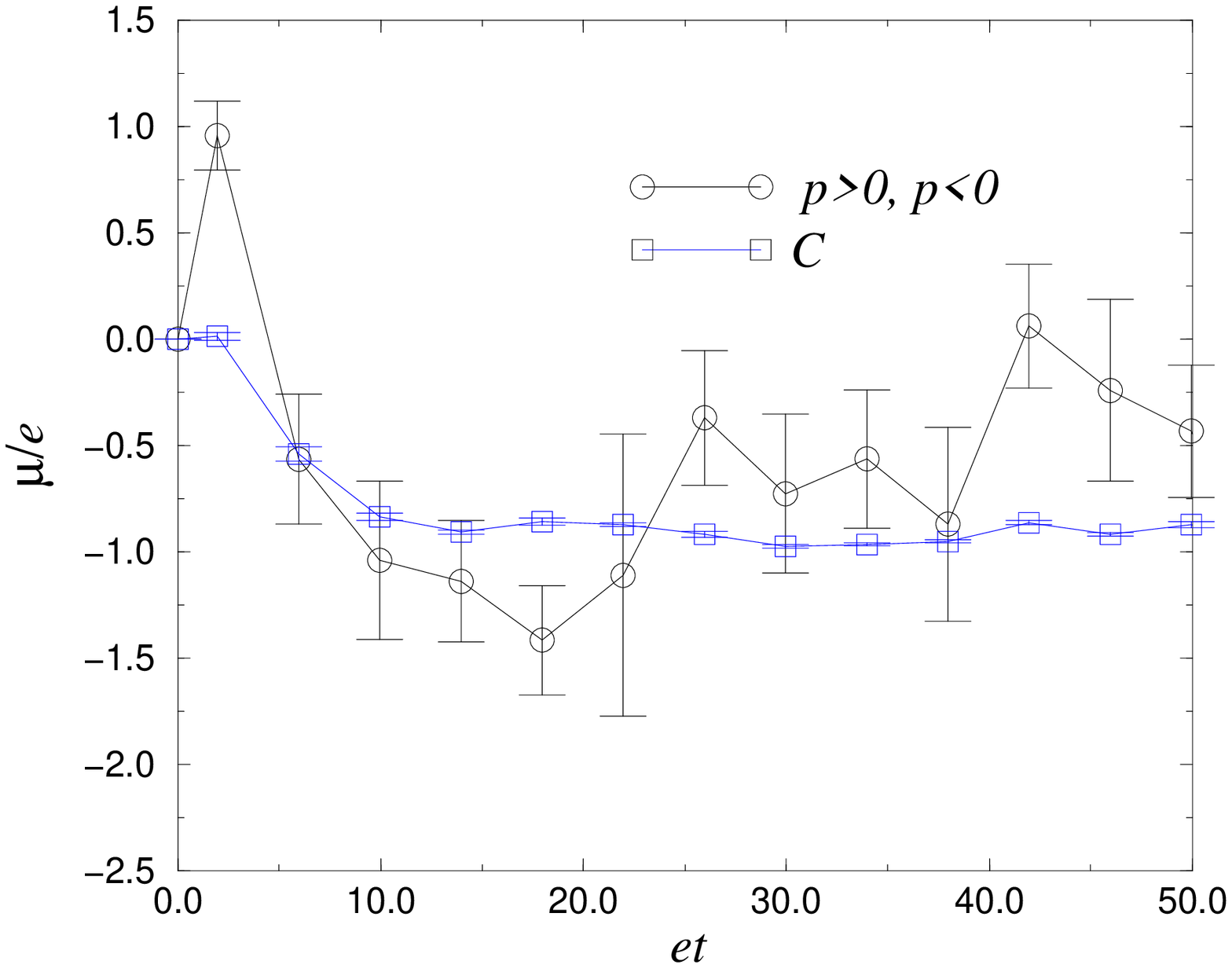,width=16.0cm}}
\caption{Time dependent effective chemical potential versus $et$, obtained 
as in the previous figure. Here the result is averaged 
over the chemical potentials obtained from the $p>0$ and $p<0$ fits. The 
line denoted with $C$ is again obtained from (\ref{eqenergy}).} 
\label{figmu}
\end{figure}

{}From least square fits to the data we can find the time 
dependent temperature and chemical potential, with errors. 
The effective temperature is obtained separately from $p>0$ and $p<0$ 
distribution functions and is shown in fig.\ \ref{figT}.
The effective chemical potential is shown in fig.\ \ref{figmu}. In this 
case we present the average of the chemical potential obtained from the 
$p>0$ and $p<0$ distribution functions, since the fluctuations
are rather large.
The errors correspond to the least squares fit.

If the fermions are in thermal and chemical equilibrium, with possibly {\em 
slowly} varying time dependent temperature and chemical potential, 
and interactions are neglected,
the relation between $\bra Q_5\ket$, $E_F$, $T$ and $\mu$ is as follows. 
For massless fermions, $E_p=|p|$, we find that in that case the axial 
charge is given by 
\[ \frac{\bra Q_5\ket}{L} = n_f\int_{-\infty}^{\infty} \frac{dp}{2\pi} 
q_{5p} \left[ \frac{1}{e^{\bt(E_p-\mu q_{5p})}+1} - 
\frac{1}{e^{\bt(E_p+\mu q_{5p})}+1} \right]\\
= n_f\frac{\mu}{\pi}.
\]
With the help of the anomaly equation, $\bra Q_5\ket =n_fC$, this gives 
$\mu$ in terms of $C$.
Likewise, the renormalized fermion energy at finite temperature and 
chemical potential can be calculated, 
\be 
\label{eqenergy}
  \frac{E_F}{L}= 
n_f\left(\frac{\pi}{6}T^2+\frac{1}{2\pi}\mu^2\right),\;\;\;\;
 \mu  = \frac{\pi C}{L}.\ee
Given $C$ and $E_F$, we can predict $T$ and $\mu$ {\em assuming}\, the 
fermions are in equilibrium and neglecting interactions.
This predictions are shown in figs. \ref{figT} and \ref{figmu} as well.
(Here the error is the standard deviation of the average in
an interval $et_{\rm av}=4$, treating the data as uncorrelated.)
We see that the equilibrium temperature obtained from (\ref{eqenergy}) lies
significantly below the effective temperatures obtained from the particle
distribution functions. 
Presumably this reflects the fact that the fermions are of course not 
really free. 

The chemical potential from the fits has large
fluctuations and errors. This is possibly due to finite size effects. 
The time independent equilibrium result (\ref{eqenergy}) suggests that 
the chemical potential will be sensitive to time dependence of the 
Chern-Simons number (see fig.\ \ref{figC}). This dependence should 
decrease when the volume $L$ gets larger.

\section{Conclusion}
\setcounter{equation}{0}
\label{secconl}

We presented numerical results for fermion particle 
production in the presence of inhomogeneous time dependent Bose fields. 
The quantum field dynamics was approximated by effective mean field 
equations of motion for the Bose fields coupled to quantized 
fermion fields, represented by mode functions.
The particle number was extracted from the single-time Wigner function, 
using the free Wigner function as a guideline. For strongly coupled 
theories, such an approach would presumably not be possible. 

A comparison of the particle number with a time dependent Fermi-Dirac
distribution, i.e.\ containing a time dependent temperature and chemical
potential, showed that the produced particles are approximately in thermal
equilibrium,  already after the first few oscillations of the Bose fields.  
The effective temperature increases when more energy is transferred towards
the fermion degrees of freedom. 
Note that these results are quite different from those 
obtained with only homogeneous Bose fields, in which case
particle distributions are typically non-thermal.
Due to the inhomogeneous gauge field, the equations studied here contain 
classical (screened) Coulomb scattering of the particle-like modes,
whereas this is absent in the case of homogeneous mean fields, for which
mode coupling is severely reduced.

In this paper our main focus
was on fermion particle production. An analysis of the complete system,
e.g.\ of 
the time dependent energy distribution of the Bose
fields, remains to be done. 

This study may be relevant for
inflationary scenarios. Preheating of the universe is usually analysed by
coupling the homogeneous inflaton field to the particle mode functions,
leading to resonance band structures and highly non-thermal 
particle distribution functions (see e.g.\ references given in the 
Introduction).
The stability of these resonance bands and
the time scales of thermalization will be affected when inhomogeneities are 
taken into account.

\subsubsection*{Acknowledgments}  

We thank Bert-Jan Nauta and Henk Stoof for useful discussions. 
This work is supported by FOM.

\renewcommand{\thesection}{\Alph{section}}
\setcounter{section}{0}
\renewcommand{\theequation}{\Alph{section}.\arabic{equation}}

\section{Lattice implementation}
\setcounter{equation}{0}
\label{seclat}

We solve the effective equations numerically, using a formulation of the 
theory on a lattice in space and time. 
For details concerning especially the fermion fields we refer to 
\cite{Aarts:1998td}, here we repeat only the necessary.
We use a lattice in space with $N$ sites and lattice spacing $a$, such 
that $L=Na$. 
The index $p$ labeling the mode functions then takes a finite number of 
values 
\be 
\label{eqplat}
p=\frac{2\pi}{L}(n-\half), \;\;\;\;n\in \{-\half N+1, \ldots, \half N\},
\ee
and the total number of mode functions is given by $2n_fN=4N$.
On the lattice the dispersion relation is modified and reads for  
massless Wilson fermions \cite{Aarts:1998td}
\be
\label{eqdisplat}
E_p = \sqrt{s_p^2+m_p^2}, \;\;\;\;s_p=a^{-1}\sin ap, 
\;\;\;\;m_p=a^{-1}(1-\cos ap),
\ee
where $m_p$ is the `Wilson mass'.
The maximal fermion momentum on the lattice equals $ap = 
\pi(1-1/N) \simeq \pi$, in units of the lattice spacing. 
For such high momenta lattice artefacts are important. 
The physical modes are those with momenta that are small in lattice 
units, since then $E_p = |p| (1 + {\cal O}(a^2p^2))$. 
A rough guideline is $|ap| \lesssim 0.5$, for $ap=0.5$ 
the correction term is approximately $1\%$. The time step is denoted 
with $a_0$. 
The lattice parameters are $N=64$ and $a_0/a=0.005$. 

\section{Numerical calculation of the Wigner function}
\setcounter{equation}{0}
\label{app}

In this appendix we describe how we calculate the Wigner function 
numerically on the lattice. We also compare the single-time Wigner function 
with the gauge fixed equal-time two-point function.

While the definition of the Wigner function is manifestly gauge 
invariant, it is particularly convenient to calculate it in the 
(completely gauge fixed) Coulomb gauge. We define this 
(suppressing the time dependence) by
\be
\label{eqA1}
A_1^c = A_1(x) + \partial_1 \theta_c(x), \;\;\;\; \partial_1 A_1^c=0, 
\;\;\;\; -1 < C_c \leq 1,
\ee
where the gauge fixed Chern-Simons number is $C_c = -LA_1^c/(2\pi)$.
This completely fixes the gauge. In particular the freedom under large 
gauge transformations is  fixed by the last requirement in (\ref{eqA1}).
We recall that due the $q=\half$ charged 
fermions, $C=1$ is not gauge equivalent to $C=0$, but e.g.\ to $C=-1$.
The fermion fields transform accordingly $\psi(x) \to \psi_c(x) = 
\exp[iq\theta_c(x)]\psi(x)$. The parallel transporter in the Wigner 
function is in this gauge simply $U(y,x;t) = \exp \left[iqA_1^c(y-x)\right]$. 

The Wigner function, averaged over the lattice in space, becomes
\be 
\label{eqA2}
W(p,t) = 
\frac{1}{L}\sum_X \sum_z  e^{-i(p+qA_1^c)z}
  \bra\psi_c(X+\half z,t)\bar{\psi}_c(X-\half z,t)\ket.
\ee
Up to the factor $\exp \left[-iqA_1^cz\right]$, this is identical 
to what would have been obtained when started from the Coulomb 
gauge fixed propagator. 
To evaluate it, we first calculate the Fourier transform of the gauge fixed 
two-point function 
\be 
\label{eqA3}
S(p,t) = 
\sum_z e^{-ipz}\frac{1}{L}\sum_X 
  \bra\psi_c(X+\half z,t)\bar{\psi}_c(X-\half z,t)\ket,
\ee
which is the discrete Fourier transform of an antiperiodic function.
Hence we know $S(p,t)$ at a discrete set of $p$ values, namely
those given by (\ref{eqplat}). 
Using inter- and extrapolation, the value of 
$S(p,t)$ at other values of $p$ can be found as well. 
It follows from comparing (\ref{eqA2}) and (\ref{eqA3}) that the Wigner 
function is given by $W(p,t) = S(p+qA_1^c,t)$.

The gauge fixed propagator shows discontinuous behaviour when the 
Chern-Simons number crosses the boundaries set by fixing the freedom 
under large gauge transformations in (\ref{eqA1}).
We expect these discontinuities to become less important when the volume is 
increased.
On the other hand, the presence of the time dependent phase factor in 
(\ref{eqA2}) leads to a smooth behaviour of the Wigner function when the 
Chern-Simons number changes during time evolution.

Finally, the explicit expressions for the time dependent particle number 
$N_p$(t), the antiparticle number $\bar N_p(t)$ and the effective mass 
$m_p(t)$ in terms of the functions ${\cal F}_{\rm av}(p,t)$ and ${\cal 
V}_{\mu\rm av}(p,t)$ can be obtained as follows. First note from 
(\ref{eqF}) and (\ref{eqV1}) that 
\[  {\cal F}_{\rm av}^2(p,t)+{\cal V}_{1\rm av}^2(p,t) = 
\frac{1}{4}\left(1-N_p-\bar N_{-p}\right)^2,
\]
which gives $|1-N_p-\bar N_{-p}|$. The sign can be found from 
(\ref{eqV1}), and \[ N_{{\rm av}p}(t) = \half (N_p+\bar N_{-p}) = 
\mbox{sign}[p{\cal V}_{1\rm av}(p,t)] 
\sqrt{ {\cal F}_{\rm av}^2(p,t)+{\cal V}_{1\rm av}^2(p,t)} 
+\half.\]
The individual (anti)particle numbers (instead of the sum) can be found by 
adding or subtracting 
\[  -{\cal V}_{0\rm av}(p,t)+\half = \half(N_p - \bar N_{-p}).\]
The effective mass, or more generally the dispersion relation, can then 
be found from 
\[ \sqrt{m_p^2(t)+p^2} = \frac{p\left( N_{{\rm av}p}(t)-\half\right)}{
{\cal V}_{1\rm av}(p,t)},\]
provided ${\cal V}_{1\rm av}(p,t) \neq 0$ of course. 
However,  ${\cal V}_{1\rm av}(p,t)$ vanishes only when 
$N_{{\rm av}p}(t)-\half$ equals zero as well, since the momentum $p$ on 
the lattice is always different from zero, due to the antiperiodic 
boundary conditions in space (see (\ref{eqplat})).    
Note that since ${\cal F}_{\rm av}(p,t)$ and ${\cal V}_{1\rm av}(p,t)$ are 
both real, 
${\cal F}_{\rm av}(p,t)/{\cal V}_{1\rm av}(p,t) = - m_p(t)/p$
is real as well,  which implies that $m_p(t)$ is real.

\newpage
\hspace{0cm}
 
\hspace{0cm}

\hspace{0cm}

\hspace{0cm}

\hspace{0cm}

\hspace{0cm}

\begin{thebibliography}{99}

\bibitem{Cooper:1994hr}
F.~Cooper, S.~Habib, Y.~Kluger, E.~Mottola, J.~P. Paz and P.~R. Anderson,
{\em Phys. Rev.} {\bf D50} (1994) 2848.

\bibitem{Kluger:1991ib}
Y.~Kluger, J.~M. Eisenberg, B.~Svetitsky, F.~Cooper and E.~Mottola, 
{\em Phys. Rev. Lett.} {\bf 67} (1991) 2427;
F.~Cooper, S.~Habib, Y.~Kluger and E.~Mottola,
{\em Phys. Rev.} {\bf D55} (1997) 6471.

\bibitem{Bo:1996}
D.~Boyanovsky, H.~J. de~Vega, R.~Holman and J.~F.~J. Salgado, 
{\em Phys. Rev.} {\bf D54} (1996) 7570;
D.~Boyanovsky, D.~Cormier, H.~de~Vega, R.~Holman, A.~Singh  and M.~Srednicki,
  {\em Phys. Rev.} {\bf D56} (1997) 1939.

\bibitem{Bo:1998}
D.~Boyanovsky, C.~Destri, H.~J. de~Vega, R.~Holman and J.~Salgado,
{\em Phys. Rev.} {\bf D57} (1998) 7388.

\bibitem{fermionswith1}
Y.~Kluger, J.~M. Eisenberg, B.~Svetitsky, F.~Cooper and E.~Mottola, 
{\em Phys. Rev.} {\bf D45} (1992) 4659.

\bibitem{fermionswith2}
D.~Boyanovsky, M.~D'Attanasio, H.~J. de~Vega, R.~Holman and D.~S. Lee,
{\em Phys. Rev.} {\bf D52} (1995) 6805;
J.~Baacke, K.~Heitmann and C.~P{\"a}tzold, 
{\em Phys. Rev.} {\bf D58} (1998) 125013.

\bibitem{Aarts:1998td}
G.~Aarts and J.~Smit,  
{\em Nucl. Phys.} {\bf B555} (1999) 355.

\bibitem{Bettencourt:1998nf}
L.~M.~A. Bettencourt and C.~Wetterich, 
{\em Phys. Lett.} {\bf B430} (1998) 140.

\bibitem{Grigorev:1988bd}
D.~Y. Grigoriev and V.~A. Rubakov, {\em Nucl. Phys.} {\bf B299} (1988) 67.

\bibitem{Khlebnikov:1996mc}
S.~Y. Khlebnikov and I.~I. Tkachev,
{\em Phys. Rev. Lett.} {\bf 77} (1996) 219.

\bibitem{baryo}
For recent reviews, see e.g.,
V.~A. Rubakov and M.~E. Shaposhnikov, 
{\em  Usp. Fiz. Nauk} {\bf 166} (1996) 493;
M.~Trodden, {\tt hep-ph/9803479}.

\bibitem{Garcia-Bellido:1999sv}
J.~Garc\'{\i}a-Bellido, D.~Grigoriev, A.~Kusenko and M.~Shaposhnikov,
{\tt hep-ph/9902449}.

\bibitem{fermionswithout}
P.~B. Greene and L.~Kofman, {\em Phys. Lett.} {\bf B448} (1999) 6;
G.~F. Giudice, M.~Peloso, A.~Riotto and I.~Tkachev, 
{\em JHEP} {\bf 08} (1999) 014.

\bibitem{Kofman:1997yn}
L.~Kofman, A.~Linde and A.~A. Starobinsky, 
{\em Phys. Rev.} {\bf D56} (1997) 3258.

\bibitem{Vasak:1987um}
D.~Vasak, M.~Gyulassy and H.~T. Elze,
{\em Ann. Phys.} {\bf 173} (1987) 462.

\bibitem{Blaizot}
J.~P. Blaizot and E.~Iancu, 
{\em Nucl. Phys.} {\bf B390} (1993) 589;
{\em Phys. Rev. Lett.} {\bf 70} (1993)  3376;
{\em Nucl. Phys.} {\bf B417} (1994) 608.

\bibitem{singletime}
P.~Zhuang and U.~Heinz, {\em Ann. Phys.} {\bf 245} (1996) 311;
S.~Ochs and U.~Heinz, {\em Ann. Phys.} {\bf 266} (1998) 351.

\bibitem{Ibaceta:1998}
D.~Ibaceta and E.~Calzetta, {\tt hep-ph/9810301}.

\end{thebibliography}
\end{document}